# The study of the efficiency of nitrogen to NV-center conversion in high nitrogen content samples

S.V. Bolshedvorskii[1,2], S.A. Tarelkin[3], V.V. Soshenko[1,2], I.S. Cojocaru[1,2,4], O.R. Rubinas[1,2,4], V.N. Sorokin[1,2], V.G. Vins[5], A.N. Smolyaninov[2], S.G. Buga[3], A.S. Galkin[3], T.E. Drozdova[3], M.S. Kuznetsov[3], S.A. Nosukhin[3], A.V. Akimov[1,2,4,6]

[1]*P.N. Lebedev Institute RAS, Leninsky prospekt 53, Moscow, 119991, Russia.*

[2]*LLC Sensor Spin Technologies, 121205 Nobel str. 9, Moscow, Russia*

[3]*Technological Institute for Superhard and Novel Carbon Materials, 7a Tsentralnaya str., Troitsk, Moscow, Russia*

[4]*Russian Quantum Center, Business center "Ural", 100A Novaya str., Skolkovo, Moscow, 143025, Russia*

[5]*LLC Velman, 630060 Zelenaya gorka str.1/3, Novosibirsk, Russia*

[6]*National University of Science and Technology MISIS, Leninsky prospekt 4, Moscow, 119049, Russia*

email: a.akimov@rqc.ru

## I.     ABSTRACT

The nitrogen-vacancy color center in diamond is one of the most important systems in the fast-growing field of sensing. This color centers are used in both high-resolution and high-sensitivity sensors. However, techniques for quick and efficient formations of this color center are still in the development stage. In this paper, we present a study on the influence of the electron irradiation dose on the conversion of substitutional nitrogen into $NV^-$ centers. The study was done on diamonds that were highly enriched with nitrogen (~100 ppm), which on one hand should maximize the effect of irradiation, and on another be of interest for high-sensitivity magnetometers. The maximum achieved conversion efficiency was as high as $37 \pm 3.7\ \%$, with no observed saturation on the electron dose even with the simplest annealing procedure. The measurements of the corresponding dephasing time made it possible to estimate for shot-noise limited sensitivity per unit volume of a stationary field sensor with such a diamond to be $9 \pm 1 \times 10^{-14} T / \sqrt{Hz \cdot mm^{-3}}$.

## II. INTRODUCTION

Color centers and other impurity-related defects are attracting a lot of attention in the last decades. In particular, negatively charged nitrogen-vacancy color center in diamond ($NV^-$-center) is a subject of active research due to a large number of potential applications, including solid-state single photon sources [1], qubits [2] and qubit registers [3,4], quantum communication lines [5,6], quantum teleportation [7] and various sensing applications [8]. Sensing applications seem to be the closest ones to practical applications and at the moment cover high-resolution thermometry [9], measurements of electric [10] and magnetic [11] fields (either locally [12] or with high sensitivity [13]), magneto resonance tomography [14], sensing of strain [15–17], rotation [18] and acceleration [19].

The high sensitivity number of utilized color centers generally improves the sensitivity of the center. Nevertheless, the large use of $NV^-$-centers faces a number of difficulties; such as, an increase in the concentration of the color centers unavoidably leads to the decrease of the diamond coherence properties, while the increase in the sample size of the diamond makes it difficult to drive color centers with high laser and microwave radiation, as well as to retain its drive uniform [20,21]. Thus, a lot of effort lately were put into growth optimization and post-growing processing of diamonds to maximize the number and efficiency of $\text{NV}^-$-centers in various sensing applications [15,22–25].

There are a number of post-growing diamond processing techniques. The most popular ones are the implantations of He ions [26,27] or electrons [28,29] with subsequent annealing. This still depends on optimal annealing temperatures and implantation doses, which seem to rely strongly on the characteristics of the sample. In particular, it has been demonstrated, that for high-sensitivity magnetometry [30] diamond plates with relatively high concentrations (~100 ppm) of nitrogen tend to have an advantage over relatively low-concentrated ones. These types of samples tend to have larger nitrogen to $\text{NV}^-$-center conversion and less unwanted absorption, and still have long enough coherence times for magnetometry.

In this paper, we focus on the optimization of diamond post-growing procedures, such as electron irradiation and annealing, and try to optimize the yield of the $\text{NV}^-$-center, as well as shot-noise limited sensitivity per unit volume of the crystal for static field (DC) magnetometry. In monocrystal diamonds, with about 100 ppm concentration of nitrogen, we observed a continuous increase in the conversion efficiency with electron dose irradiation, limited by the time available at the electron-irradiation machine.

### III. METHODS

The $\text{NV}^-$-center is formed by a combination of nitrogen atoms, substitution carbon in a diamond lattice, and the vacancy next to it. Interesting is the negatively charged state of this color center, with extra electrons captured by the center. Energy levels (Figure 1a) of the color center are located in a diamond bandgap and from triplet grounds and excited states. Besides, there is an optically inactive singlet state, which is responsible for different fluorescence intensities for levels of excitation with different magnitudes of the magnetic quantum numbers. Thus, if components of the $NV^-$-center ground state are mixed with microwave radiation, resonant to the $|m_s=0\rangle \leftrightarrow |m_s=+1\rangle$ transition (see Figure 1a), the fluorescence of the color center is lower than if the microwave radiation is absent or off-resonance. The resonance in the fluorescence spectrum with respect to microwave radiation frequency is typically called optically detected magnetic resonance and is the key feature enabling the use of the $NV^-$-center.

The diamond samples (type Ib), enriched with nitrogen, were grown using a temperature-gradient, high temperature, high pressure (TG-HPHT) growth process [31,32] at a pressure of 5 GPa and a temperature of 1500-1550°C. As a material for growing pure (99,9995%) graphite was used. Alloy Fe-C was used as a solvent, while nitrogen was not artificially added. The residual air in the growing mix served as a rich enough source of nitrogen.

The diamond crystals were then cut into plates with a thickness of $300\ \mu m$. The plates were cut from the growing sector {111} with crystallographic base surface orientation (100) and were measured to contain substitutional nitrogen (C-centers) concentration of about 1.5×10$^{19}$ cm$^{-3}$. The cutting was performed using a home-made laser cutter based on InnoLas Laser (Nanio 532-18-y) and then mechanically polished so that the average absolute values of the profile height deviations ($R_a$ coefficient of roughness) became smaller than 5 nm. The concentration of C-center was measured using an infrared absorption method and a Fourier spectrometer Bruker Vertex 80v, combined with an optical microscope HYPERION 2000.

In contrast to the CVD-grown crystal, TG-HPHT monocrystalline has practically no vacancies after growth. Thus, the implantation of the vacancies is an essential step for the formation of the $NV^-$

-centers. In this work, we utilize electron-implantation using linear electron accelerators with energy up to 2 MeV (LLC Velman, Novosibirsk) and 3.5 MeV INDUSTRIAC 3E1000M (TISNCM). In both cases, the energy of the electrons was chosen to be 2 MeV, while doses of irradiation $d$ were varied in range of $1\times10^{17} - 5\times10^{18}\,\text{cm}^{-2}$. The maximum dose was limited by the time, available at the electron accelerator. The irradiation was followed up by vacuum annealing at a temperature of 700-800 Cº for 2 hours.

Concentrations of the formed after annealing $\text{NV}^-$-centers were measured using the visible absorption spectrum (see Figure 1b), taken at 77 K using the method described in references [31,32].

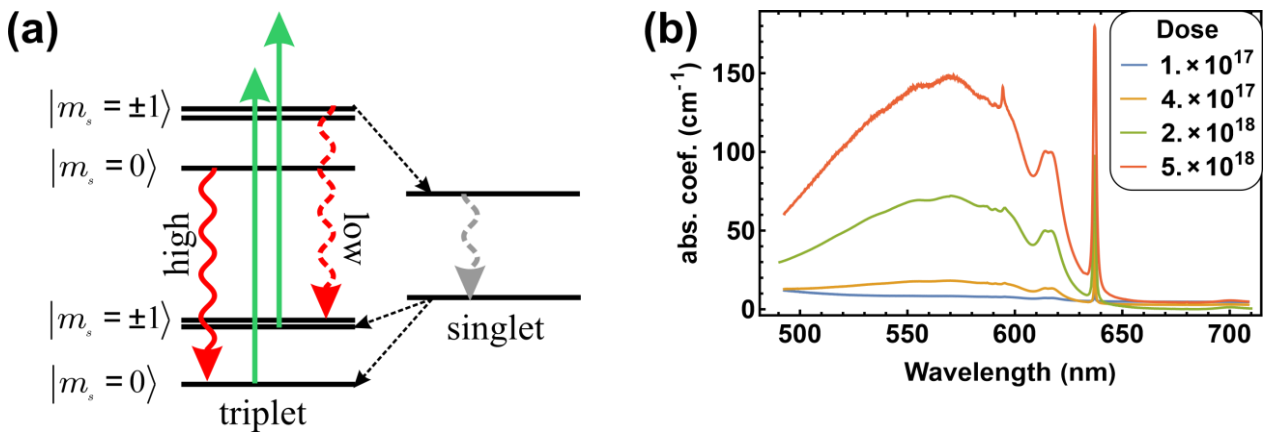

Figure 1 a) $\text{NV}^-$-center simplified level scheme. b) Absorption spectra diamond plates at 77 K

Dephasing time $T_2^*$ of ensembles of $\text{NV}^-$-center was performed using a home-made confocal microscope (similar to the one described in [22]). To manipulate the spin state of the $\text{NV}^-$-centers microwave field was supplied to the sample. The field was radiated by a Helmholtz resonance antenna [33], powered by an amplifier (Minicircuits ZHL-16W-43+) which was fed by a Stanford Research Systems SG384 generator.

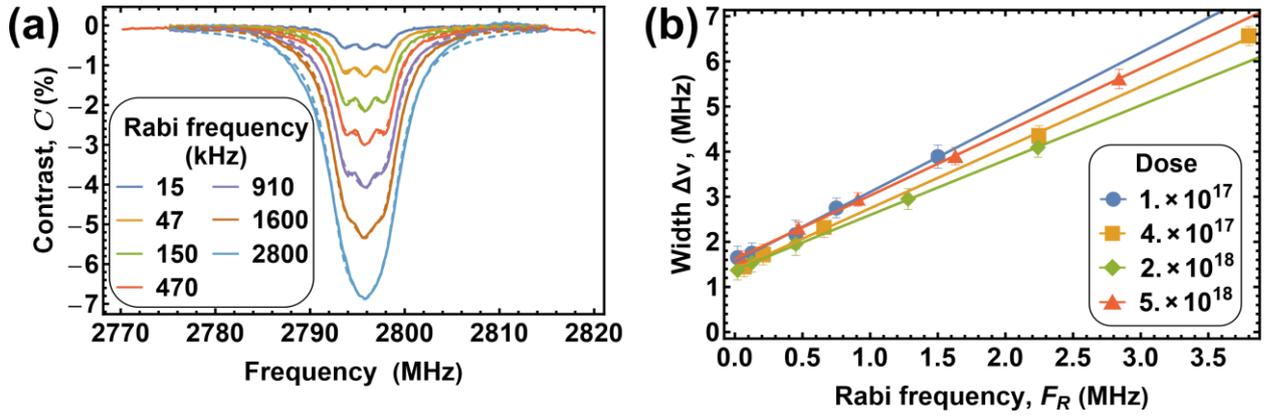

Figure 2 a) ODMR spectra of the ensemble of NV⁻-centers for a diamond plate with a dose of $5 \times 10^{18}$ cm$^{-2}$ versus applied microwave power. The Rabi frequency of the microwave transition varied in the 0.015-2.8 MHz range. b) Dependence of the ODMR resonance peak width $\Delta \nu$ on the Rabi frequency of the microwave transition for all samples used.

The dephasing time $T_2^*$ was extracted from the dependence on the Rabi frequency of the ODMR width (see Figure 2a). The ODMR spectra were recorded at different microwave powers and then fitted with a combination of three Lorentzian functions with fixed detuning between the function maximums of 2.2 MHz, corresponding to the well-known ODMR splitting line, due to the interaction of the NV⁻ electron spin with nuclear spins of the nitrogen atom within the color center [34]. Rabi frequencies were measured using a standard technique; the centers were first initialized using green laser excitation, and then the microwave pulse of variable length was applied [30]. The microwave frequency was set to the central frequency of the ODMR line at 2795 MHz (see Figure 2a). The fluorescence of color centers was then measured in a 500 ns time window right after the microwave pulse. The resulting graph of fluorescence versus the length of the microwave pulse was then fitted with a product of exponential decay envelop and sin-shape. The fitted frequency of the sin-shape was then used as a Rabi frequency $F_R$.

The power dependence of the extracted full width at half maximum (FWHM) $\Delta \nu$ was then approximated by a straight line and the value of the width at 0 power $\Delta \nu$ was extracted (see Figure 2b). The dephasing time $T_2^*$ was then calculated [35] as:

$$T_2^* = \frac{1}{\pi \cdot \Delta \nu} \qquad (1)$$

## IV. RESULTS AND DISCUSSIONS

A summary of the measurement results of the nitrogen concentration, concentration of $NV^-$ and the corresponding dephasing time for various irradiation doses, is presented in Table 1. The table also contains the conversion efficiency for the substitutional nitrogen into $NV^-$-centers, which was estimated using the following formula:

$$\gamma = \frac{n_{NV}^*}{n_C}, \qquad (2)$$

where $n_{NV}^*$ in the nitrogen concentration after the annealing procedure and $n_C$ is substitutional nitrogen concentration before diamond irradiation

Table 1 Summary of the measurements.

| # | $n_C$, cm$^{-3}$ | $d$, cm$^{-2}$ | $n_{NV}^*$, cm$^{-3}$ | $\gamma$, % | $T_2^*$ | $\eta_{DC}$, T$/\sqrt{Hz \cdot mm^{-3}}$ |
|---|---|---|---|---|---|---|
| 1 | $1.5 \times 10^{19}$ | $1 \times 10^{17}$ | $3 \pm 0.3 \times 10^{16}$ | $0.2 \pm 0.02$ | $231 \pm 12$ | $1.09 \pm 0.12 \times 10^{-12}$ |
| 2 | $1.5 \times 10^{19}$ | $4 \times 10^{17}$ | $1.4 \pm 0.15 \times 10^{17}$ | $0.93 \pm 0.1$ | $228 \pm 10$ | $5.2 \pm 0.6 \times 10^{-13}$ |
| 3 | $1.5 \times 10^{19}$ | $2 \times 10^{18}$ | $2.3 \pm 0.23 \times 10^{18}$ | $15.3 \pm 1.5$ | $204 \pm 10$ | $1.35 \pm 0.15 \times 10^{-13}$ |
| 4 | $1.5 \times 10^{19}$ | $5 \times 10^{18}$ | $5.6 \pm 0.56 \times 10^{18}$ | $37 \pm 3.7$ | $197 \pm 9$ | $9 \pm 1 \times 10^{-14}$ |

Figure 3 illustrates a correlation between irradiation dose and conversion of substitutional nitrogen centers into the $NV^-$-centers. The figure suggests a linear dependence of the conversion efficiency with no saturation observed. We should stress, that the observed conversion efficiency of $37\pm4\%$ exceeds many recently published results, including detailed studies and optimizations of the annealing procedure, while the annealing procedure in this work is rather trivial and is only the first step of much more advanced ones. This suggests that the irradiation dose for the high-concentrated samples has a very large impact on the final conversion efficiency and must be kept as high as possible. We could expect that the application of a more optimal annealing procedure may further enhance the conversion efficiency.

Also, despite rising conversion efficiency with the irradiation dose, we observe a slight drop in the dephasing time. Increasing the irradiation dose is expected to increase the conversion of substitutional nitrogen into NV-centers, as observed. The dephasing time on NV-center with 100 ppm concentration is typically limited by the concentration of the substitutional nitrogen, and thus expected to increase as the nitrogen concentration decreases. But, experimentally the decrease in the dephasing time indicated that the NV-NV interaction started to limit the dephasing time and further confirm a high conversion efficiency for NV-centers [1].

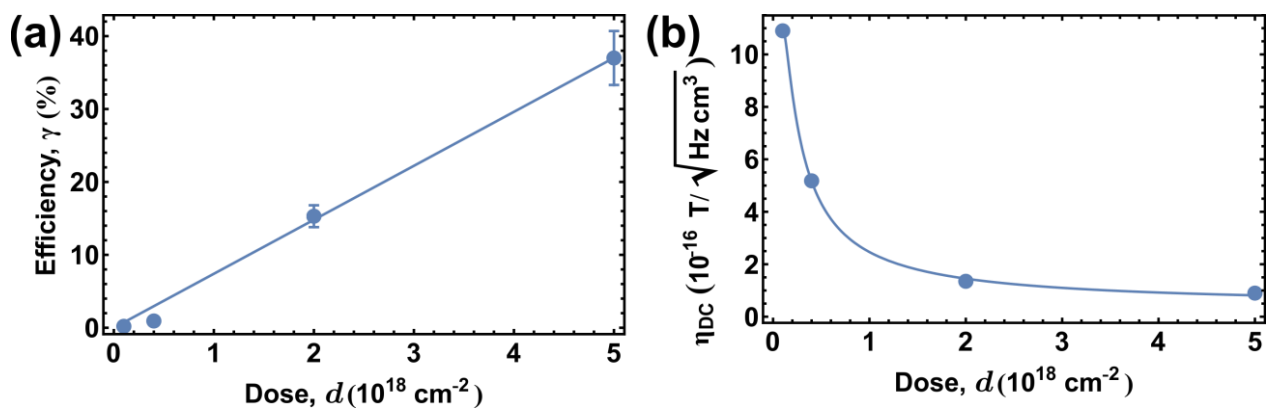

Figure 3. a) Correlation between C-centers to $NV^-$-centers conversion efficiency and irradiation dose. b) Shot-noise limited sensitivity of a DC magnetometer versus

irradiation dose. The line represents an eye-guiding fit function $0.38 + 2.2x^{-1} - 0.11x^{-2}$.

The concentration of the color centers and the efficiency of substitutional nitrogen to $NV^-$-centers are not the only parameters which affect the sensitivity of a sensor. The other important parameter is the dephasing time, which allows a shot-noise limited estimate of potential sensitivity for DC magnetometry per unit volume of diamond [11]:

$$\eta_{DC} = \frac{\hbar}{g_e \mu_B} \frac{1}{C\sqrt{nT_2^*}} \qquad (3)$$

Calculated this way $\eta_{DC}$ are summarized in Table 1 and presented in Figure 3b. It could be clearly seen that the sensitivity improves with dose even with a relatively simple annealing procedure. More advanced annealing procedures have the potential to further improve dephasing time and sensitivity. Moreover, the increased conversion efficiency from substitutional nitrogen to $NV^-$-centers solves one more issue – unwanted absorption of the diamond sample by C-centers, thus improving sensitivity per laser power [30].

While this research only was done for one specific concentration on substitutional nitrogen centers, it is likely, that a considerable increase in electron irradiation dose may help samples with different, even with a relatively low concentration of nitrogen. One can speculate, that in the case of lower nitrogen concentrations the irradiation dose should be even larger, since lower nitrogen concentration reduces the probability of nitrogen-vacancy collision on an annealing step. Nevertheless, previously some saturation of conversion efficiency was observed for samples with low nitrogen dose [36], thus further research is necessary to confirm this hypothesis.

## V. SUMMARY

The influence of the electron irradiation dose on the conversion of substitutional nitrogen into $NV^-$-centers was investigated on an example of diamond plates with about 100 ppm of nitrogen ($1.5 \times 10^{19} \, cm^{-3}$). A strong linear correlation between the dose of high-energy electron (2 MeV) irradiation and conversion efficiency of substitutional nitrogen to $NV^-$-centers was found. No saturation was detected in the rage of irradiation doses of $10^{17} - 5 \times 10^{18} \, cm^{-2}$, the conversion efficiency was likely restricted by the available time on the electron accelerator. The annealing step was not optimized, and a simple one-step 800 C annealing was used. The maximum achieved conversion efficiency was as high as $37 \pm 3.7$ %. The dephasing times in the $NV^-$-centers were measured for the same irradiation doses. The decrease in the dephasing time with the increase in the dose was found to be insignificant. The projected sort-noise limited sensitivity per unit volume of DC magnetometer for the plates thus was improved with irradiation dose and the best possible sensitivity with diamonds used was estimated as $9 \pm 1 \times 10^{-14} \, T / \sqrt{Hz \cdot mm^{-3}}$

## VI. ACKNOWLEDGMENTS

This work was supported by Russian Science Foundation, grant # 21-42-04407. Equipment of Common use Facility of Technological Institute for Superhard and Novel Carbon Materials(http://tisnum.ru/suec) was used.